\documentclass[preprint]{epl}

\title{Numerical indications of a $q$-generalised central limit theorem}
\author{Luis G. Moyano\inst{1,2} \and Constantino Tsallis\inst{1,2} \and Murray Gell-Mann\inst{2}}
\institute{
  \inst{1} Centro Brasileiro de Pesquisas F\'\i sicas, Rua Xavier Sigaud 150,
  22290-180 Rio de Janeiro-RJ, Brazil\\
  \inst{2} Santa Fe Institute, 1399 Hyde Park Road, Santa Fe, New Mexico 87501,  USA.  
}

\begin{document}

\maketitle

\begin{abstract}
We provide numerical indications of the $q$-generalised central limit theorem that has been conjectured (Tsallis 2004)  in nonextensive statistical mechanics. We focus on $N$ binary random variables
correlated in a {\it scale-invariant} way. The correlations are introduced by imposing the Leibnitz rule on a probability set based on the so-called $q$-product with $q \le 1$.  We show that, in the large $N$ limit  (and after appropriate centering, rescaling, and symmetrisation), the emerging distributions are $q_e$-Gaussians, i.e., $p(x) \propto [1-(1-q_e)\, \beta(N) x^2]^{1/(1-q_e)}$, with  $q_e=2-\frac{1}{q}$, and with coefficients $\beta(N)$ approaching finite values $\beta(\infty)$. The particular case $q=q_e=1$ recovers the celebrated de Moivre-Laplace theorem.

\end{abstract}

\section{Introduction}

The central limit theorem (CLT) is a cornerstone of probability theory
and is of fundamental importance in statistical mechanics. This important theorem implies, roughly speaking, that any
sum of $N$ \emph{independent} random variables will tend, as $N \to\infty$, to be distributed according
to a certain law (which behaves as an attractor in the space of distributions). When the distribution of the individual random variables has any {\it finite} variance,  the attractor for the sum will be a normal (Gaussian) distribution~\cite{b.CLT}, and this is the result
usually known as CLT (from now on denoted $G$-CLT). 
Several extensions of the CLT exist, such as the one due to Gnedenko,
Kolmogorov, and L\'evy~\cite{b.CLT} (from now on denoted $L$-CLT), widely known in physics because of its relation
with anomalous diffusion~\cite{b.anomalous}. This
extension states that the sum of \emph{independent} {\it infinite}-variance
variables will be attracted to L\'evy  distributions. 
The $G$-CLT explains the frequent occurrence of normal distributions in nature. Its first manifestation in mathematics was due to Abraham de Moivre in 1733, followed independently by Pierre-Simon de Laplace in 1774. The distribution was rediscovered by Robert Adrain in 1808, and then finally by Carl Friedrich Gauss, who based on it his famous theory of errors \cite{b.Stigler}. A central result is the fact that the binomial distribution approaches, for $N\to\infty$ and after being appropriately centralised and rescaled, a Gaussian. This can be considered as the first historical manifestation of the $G$-CLT. It is frequently referred to as the {\it de Moivre-Laplace theorem}. It is this relation that we aim to generalise here by allowing for the presence of scale-invariant global correlations (previous attempts along similar lines are reviewed in \cite{b.jonalasiniosornette}). We thus suggest an explanation of the frequent occurrence of $q$-Gaussians in natural and artificial systems \cite{b.frequent,b.tsallis}.  The basic statistical-mechanical program is still essentially Boltzmann's program in fact, until now only partially fulfilled despite a common belief to the contrary. It consists of (i) constructing, from microscopic dynamics, the probabilities of occupancy of phase space for a given (typically large) time $t$ and a given (typically large) number of elements $N$, and (ii) deriving, from these probabilities, the attractor in distribution space, the entropy, and all other thermodynamical quantities. The present paper addresses a relevant aspect of the second step only, namely the $N \to\infty$ limit for fixed (typically large) $t$.

The $q$-Gaussians are distributions that naturally emerge within the framework of ``nonextensive statistical mechanics"~\cite{b.tsallis}. They are defined by $p(x) \propto 
e_{q}^{-\beta \,x^2}\equiv
[1-(1-q)\, \beta\, x^2]^{1/(1-q)}  $, where $\beta$ is a positive constant characterising the width. They optimise 
\footnote{To be more precise, they maximise (minimise) $S_q$ whenever it is a concave (convex) function, i.e., for $q>0$ (for $q<0$). Let us also mention that, for $q<0$, only states with nonzero probability enter into the calculation of $S_q$.} 
the entropy $S_q \equiv \frac{1-\int dx [p(x)]^q}{q-1}$ (with $S_1 = S_{BG} \equiv -\int dx p(x) \ln p(x)$, where $BG$ stands for {\it Boltzmann-Gibbs}) under simple constraints \cite{b.Prato}. It has a compact support for $q<1$, recovers the Gaussian distribution for $q=1$, and decays asymptotically as a power law for $1<q<3$; $p(x)$ is not normalizable for $q \ge 3$.  Its variance $\int_{-\infty}^{\infty} dx \,x^2 p(x)$ is finite for $q < 5/3$, and diverges for $5/3 \le q \le 3$.  Its $q$-variance $\int_{-\infty}^{\infty} dx\, x^2 [p(x)]^q/ \int_{-\infty}^{\infty} dx\,  [p(x)]^q$ remains finite for $q<3$. 
It recovers the $t$-Student distribution with $l$ degrees of freedom if $q=(3+l)/(1+l)$. For $l=1$, hence $q=2$, we get the Cauchy-Lorentz distribution. 

The frequent occurrence of these $q$-distributions can be easily understood if some new CLT (from now on denoted $q$-CLT) exists. The already known theorems do not explain this quasi-ubiquity. Indeed, the convolution of $N$ {\it independent} such distributions leads, for $N\to\infty$, to Gaussians if $q<5/3$, and to L\'evy distributions if $5/3<q<3$. Therefore, these $N$ variables must be strongly correlated for the $q$-distributions to be stable under convolution, i.e., to constitute attractors in the space of distributions. In other words, a new theorem would be very welcome. 
Such a possibility was already discussed in \cite{b.Bologna} and recently conjectured in detail \cite{b.Milano}. In the light of the arguments in \cite{b.TGS}, it seems natural to think that   strictly or asymptotically scale-invariant   correlations will yield a suitable $q$-CLT.  We have not shown so far that it is so,  but we present here a $q$-generalisation of an important manifestation of the $G$-CLT, namely the de Moivre-Laplace theorem.

\section{Model}

Let us consider the simple case of $N$ identical and distinguishable binary
random variables. These variables are \emph{not} necessarily independent, and we  denote by $r_{N,n}$ the associated probabilities. We have $N$ sets of probabilities with $(N+1)$ elements each, and $n=0, 1, 2, \ldots, N$ as the variable
index within each set. We construct these sets with a special correlation relating the $(N+1)$-set to the $N$-set,  in such a way that the system has a particular \emph{scale invariance}. The probabilities are correlated across different system sizes, the {\it marginal} probabilities of the $N$-system being identical to the {\it joint} probabilities of the $(N-1)$-system. More particularly, we impose the {\it Leibnitz rule}, soon to be defined.

The trivial case is that of independence. Consider the {\it Pascal triangle},  a number triangle whose rows are
formed by the binomial coefficients ${N \choose n}=\frac{N!}{(N-n)! \,  n!}$. The set $\{ {N \choose n}     /2^N\}$ constitutes a probability set for any fixed $N$. In the limit of 
$N\rightarrow\infty$ and after appropriate centralisation and rescaling, this set approaches a Gaussian distribution. As mentioned earlier, this is known as the de Moivre-Laplace theorem. If each one of the binary variables has probabilities $p$ and $1-p$, the elements of this triangle for fixed $N$ will be given by $\{{N \choose n}p^{N-n}(1-p)^n \}$. The previous simple case (Pascal triangle) corresponds to $p=1/2$. 
As for one of the systems studied in \cite{b.TGS}, we now construct our probabilities by imposing the following rule:
\begin{equation}
\label{e.Leibniz}
r_{N,n}+r_{N,n+1} =r_{N-1,n}\,\;\;\;(n=0,1,...,N-1;\,N=2,3, ...).
\end{equation}

This rule, already referred to as the ``Leibnitz rule'', is the one used to build the \emph{Leibnitz Harmonic
Triangle}~\cite{b.Leibniz}. Note that
every probability from row $N-1$ is the sum of two probabilities from row
$N$. Furthermore, the Leibnitz rule ensures by construction that for any set of $N$ variables, the
sum of the probabilities (each one multiplied by the degeneracy factor given by the appropriate 
binomial coefficient) will always be equal to the corresponding sum for the previous row. This means that if the $(N-1)$-th row  sums to unity, so does the $N$-th row . We thus verify that $\sum_{n=0}^N {N \choose n}\ r_{N,n}=1 \;\;\;(r_{N,n} \in [0,1]; \, N=1,2,3,... ;\, n=0,1,...,N)$.

Within this procedure, the knowledge of all the elements in row $(N-1)$ and any element of row
$N$ completely determines the other $N$ elements of row $N$. 
Using Eq. (\ref{e.Leibniz}) we can analytically calculate all the probability elements of all rows and obtain 

\begin{equation}
\label{e.supergeneralform}
r_{N,n}=\sum^{N}_{i=N-n}(-1)^{i-N+n} {n \choose i-N+n} ~r_{i,0}\,, \,
\end{equation}
where each $r_{N,0}$ is an arbitrary probability value.

The only remaining question is how to choose  the set $\{ r_{N,0}\}$. In the case of probabilistic independence we simply have $ r_{N,0}=p \times p \times \ldots  \times p =p^N$ ($0\le p \le 1;\, N=1,2,3,...$) and thus $ r_{N,n}=p^{N-n}p^n$ ($n=0,1,2,...,N$).
The generalisation we shall propose here is based on the {\it $q$-product} \cite{b.qproduct}: 

\begin{equation}
\label{e.qproddefinition}
x \otimes_q y \equiv [x^{1-q}+y^{1-q}-1]^{1/(1-q)} \;\;\;\;(x,y \ge 1; q \le 1).
\end{equation}
This generalised product has the following properties: (i) $x \otimes_1 y=x \, y$; (ii) $x \otimes_q 1=x $; (iii) $\ln_q (x \otimes_q y) = \ln_q x + \ln_q y$, with $\ln_q x \equiv \frac{x^{1-q}-1}{1-q} $ $(\ln_1 x=\ln x)$ being the inverse of $e_q^x$; (iv) $\frac{1}{x \otimes_q y}=(\frac{1}{x}) \otimes_{2-q}( \frac{1}{y})$. If the probability distribution in phase space is uniform within a volume $W$, the entropy $S_q$ is given by $S_q=\ln_q W$. Property (iii) can then be interpreted as $S_q(A+B)=S_q(A)+S_q(B)$ where $A$ and $B$ are subsystems that are not independent but rather satisfy $W_{A+B}=W_A \otimes_q W_B$. This fact connects the present work with \cite{b.TGS}. 
The possibility of a correspondence between this $q-product$ with a $q$-CLT has already been conjectured \cite{b.Milano}, and some efforts along this line already exist in the literature~\cite{b.suyari}. 

Let us now proceed with our $q$-generalised de Moivre-Laplace theorem. We choose 

\begin{equation}
\label{e.qproduct}
(1/r_{N,0})=(1/p)\otimes_{q} (1/p)\otimes_{q} (1/p)\otimes_{q} \ldots \otimes_{q} (1/p) \,,
\end{equation}
hence
\begin{equation}
 r_{N,0}=p\otimes_{2-q} p\otimes_{2-q} p\otimes_{2-q} \ldots \otimes_{2-q} p=1/\, [Np^{\,q-1}-(N-1)]^{1/(1-q)} \,.
\end{equation}
For $0 <p<1$ we see that $r_{N,0}=p^N=e^{-N \ln(1/p)}$ if $q=1$, whereas $r_{N,0}\sim \frac{1}{[(1/p)^{1-q}-1]^{1/(1-q)}} \frac{1}{N^{1/(1-q)}}$ $\propto 1/N^{1/(1-q)}$ ($N \to\infty$) for $q<1$. 
Combining Eqs. (\ref{e.supergeneralform}) and (5), we obtain

\begin{equation}
\label{e.generalform}
r_{N,n}=\sum^{N}_{i=N-n}(-1)^{i-N+n} {n \choose i-N+n}
\frac{p}{[i-(i-1)p^{1-q}]^{\frac{1}{1-q}}}\,.  \,
\end{equation}
Note that  $(q,p)=(0,1/2)$ reduces to the usual Leibnitz triangle (i.e., $r_{N0}=1/(N+1)$) \cite{b.Leibniz}.


\section{Results}

We studied our model numerically as a function of the
index $q$ for typical values of $p$ and $N>>1$. To calculate the probability values $r_{N,n}$ from
Eq. (\ref{e.generalform}) we used an arbitrary precision library \cite{b.library} in order to overcome the effect of the alternating
series (i.e., subtraction of almost equal large numbers), whose relative
error grows very rapidly with the number of elements
$N$. For example, for $N=300$ and $N=1000$ we used respectively 150 and 500 significant decimal digits.
\begin{figure}[h]
\onefigure[width=11cm,angle=-90]{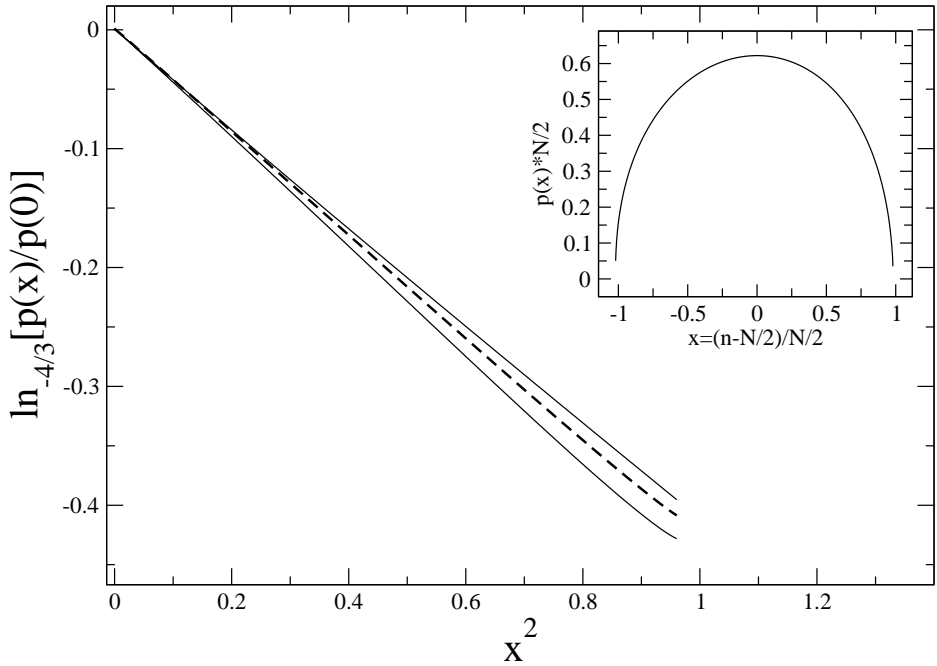}
\caption{$ \ln_{-4/3} \frac{p(x)}{p(0)}$ {\it vs} $x^2$ for $(q,p)=(3/10,1/2)$, and $N=1000$. Two branches are observed due to the asymmetry emerging from the fact that we have imposed the $q$-product on the ``left" side of the triangle; we could have done otherwise. The mean value of the two branches is indicated in dashed line. It is through this mean line that we have numerically calculated $q_e(q)$ as indicated in Fig. 3. In order to minimise the tinny asymmetry, we have represented a variable $x$ slightly displaced with regard to $\frac{n-(N/2)}{N/2}$ so that the center $x=0$ precisely coincides with the location of the maximum of $p(x)$. 
INSET: Linear-linear representation of $p(x)$.}
\label{f.qgaussian}
\end{figure}
For $p=1/2$, $N>>1$ and $q\le 1$, the probabilities 
${N \choose n}r_{N,n}$ neatly approach (see Figs. 1 and 2) the $q_e$-Gaussians
$p(x) =  A(q_e) \sqrt{\beta} \, e_{q_e}^{-\beta \, x^2}$, where
$A(q_e)$ is determined through normalization, and $x \equiv \frac{n-(N/2)}{N/2}$ is a conveniently centered and rescaled variable. 
The value of $q_e$ is obtained by plotting $\ln_{q_e}[p(x)/p(0)]$ {\it versus} $x^2$ and finding the value of $q_e$ which produces the largest linear correlation coefficient (see Fig. 1). 
This is repeated for typical pairs $(q,p)$. We see that there is some asymmetry in 
the distribution. More precisely, the $x>0$ and $x<0$ branches lead to the same $q_e$, but the corresponding slopes $\beta$ are slightly different. This asymmetry depends on $(q,p,N)$. Our main focus being the index $q_e$, we calculate the mean of both branches, and then we fit as illustrated in Fig.~\ref{f.qgaussian}.
\begin{figure}[h]
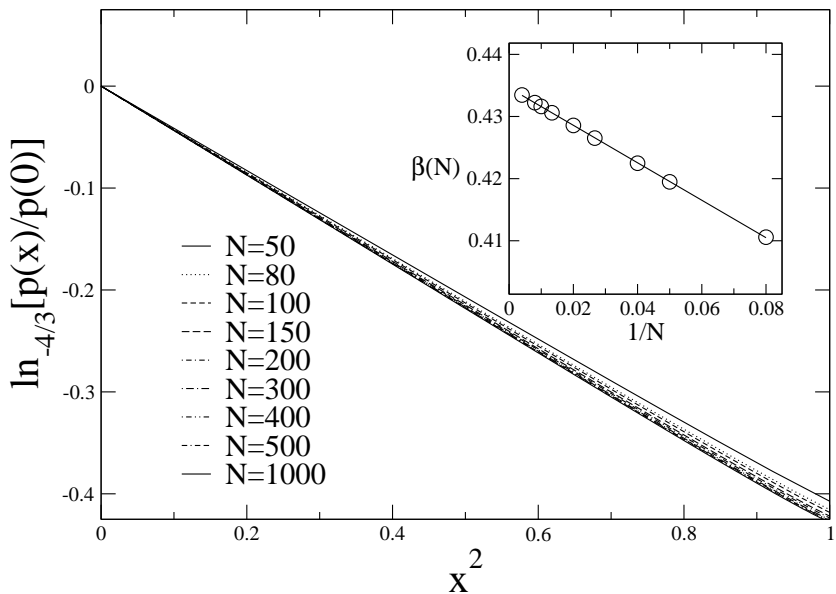

\onefigure[width=11cm]{fig2.eps}
\caption{$\ln_{-4/3} \frac{p(x)}{p(0)}$ {\it vs} $x^2$ for $(q,p)=(3/10,1/2)$ and various system sizes $N$.  INSET: $N$-dependence of the (negative) slopes of the $\ln_{q_e}$ {\it vs} $x^2$ straight lines. 
We find that, for $p=1/2$ and $N>>1$, 
$\langle(n- \langle n \rangle)^2 \rangle \sim N^2/\beta(N) \sim a(q)N + b(q)N^2$. For $q=1$ we find $a(1)=1$ and $b(1)=0$, consistent with {\it normal} diffusion as expected. For $q<1$ we find $a(q)>0$ and $b(q)>0$, thus yielding {\it ballistic} diffusion. The linear correlation factor of the $q-log \,versus\, x^2$ curves range from 0.999968 up to near 0.999971 when $N$ increases from 50 to 1000. The very slight lack of linerarity that is observed is expected to vanish in the limit $N \to\infty$, but at the present stage this remains a numerically open question.   
}
\label{f.qgaussianwN}
\end{figure}
In Fig.~\ref{f.qgaussianwN} we illustrate the dependence of the distributions on size $N$. The $q$-dependence of $q_e$ is exhibited in Fig.~\ref{f.qevsq}. The
numerical results are remarkably well described by the following conjecture:
\begin{equation}
\label{e.conjecture}
q_e=2-\frac{1}{q}\;\;\;\;\;(0 \le q \le 1).
\end{equation}
\begin{figure}[h]
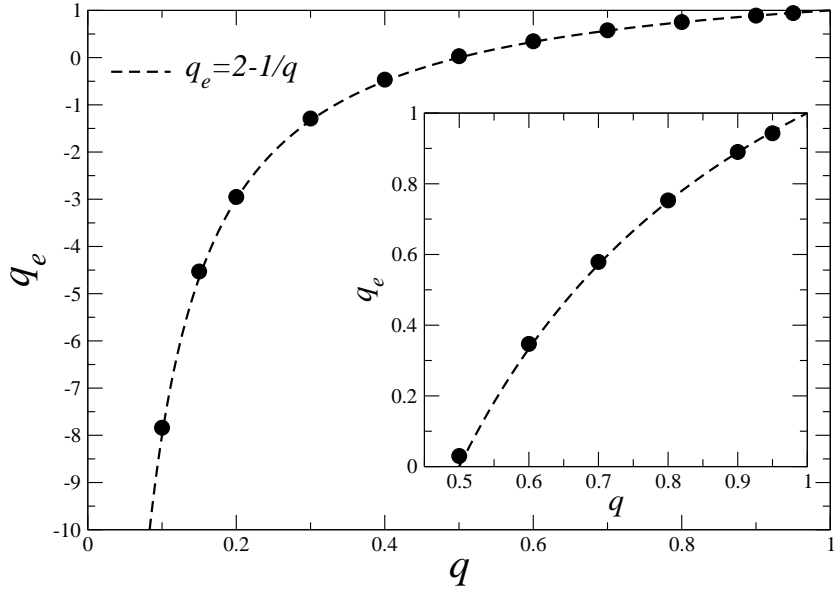

\onefigure[width=11cm]{new_qxepvsq_winset.eps}
\caption{Relation between the index $q$ from the $q$-product definition, and
 the index $q_e$ resulting from the numerically calculated probability
  distribution. The agreement with the analytical conjecture $q_e=2-\frac{1}{q}$ is
  remarkable. INSET: Detail for the range $0<q_e<1$.}
\label{f.qevsq}
\end{figure}
This of course means that we can rewrite the formula through which we introduced the global correlations (Eq. (5)) as follows: $ r_{N,0}=1/\, [Np^{\,(q_e-1)/(2-q_e)}-(N-1)]^{(2-q_e)/(1-q_e)}$, with $q_e \le 1$. If we choose this way of introducing correlations, then of course only one index is necessary within the theory, namely $q_e$, the index of the $N\to\infty$ attractor in the space of the distributions. 
We also notice that  relation (7) can be thought of as being
the composition of two \emph{dualities}, namely the {\it additive} duality $q \rightarrow (2-q)$  and the {\it multiplicative} one 
$q \rightarrow 1/q$.  
These are often encountered  (see, for instance,\cite{b.duality,b.suyari}) in the nonextensive theory 
\footnote{These two dualities appear in fact quite naturally in the theory through the properties  $\ln_q (1/x) + \ln_{2-q} x=0$ and $q \ln_q x + \ln_{1/q}(1/x^q)=0$.}. 
\begin{figure}[h]
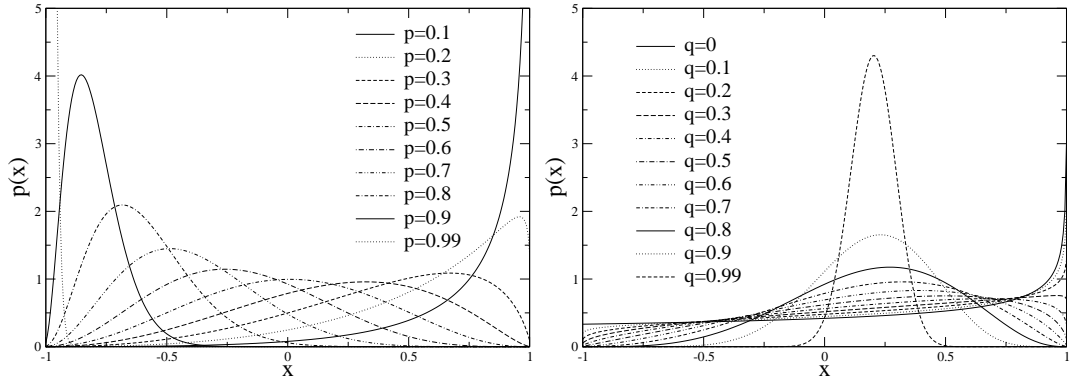

\twoimages[width=7cm]{fig4a.eps}{fig4b.eps}
\caption{Probability distribution $p(x)$ for $N=300$. {\it Left:} For $q=7/10$ and typical values of
  $p$ (the  asymmetry becomes evident for values of
  $p\neq\frac{1}{2}$). {\it Right:} For $p=4/10$ and typical values of
  $q$.}
\label{f.3}
\end{figure}

Finally, we studied the dependence of $p(x)$ on $(q,p)$: see Fig. 4. It
can be seen that the effect of varying either $p$ or $q$ is similar, namely to modify the location and height of the maximum of
the probability distribution $p(x)$, thus yielding skewness. This asymmetry reflects the particular family on which we have applied the $q$-product. Here we have done it on $r_{N0}$, i.e., on the ``left" side of the triangle. We could of course do it on its ``right" side, or on any other intermediate positions. This asymmetry is somewhat similar to the one which can occur for L\'evy distributions. A further study of the detailed influence of
these parameters is currently in progress.

\section{Summary and discussion}

We numerically illustrated, by generalising the de Moivre-Laplace theorem, the $q$-generalisation of the standard Central Limit Theorem for specially \emph{correlated} variables. The correlation is based on the $q$-product and is \emph{scale-invariant} since the Leibnitz rule has been imposed. 
Our main result is that, for the sum of $N$ 
random variables with $N>>1$, the distributions are \emph{neither} Gaussians {\it nor} L\'evy
distributions, but a different  attractor distribution which, for $p=1/2$ (and possibly other values of $p$), is a double branched $q$-Gaussian. This result strongly links the possible CLT (conjectured some time ago; see \cite{b.Milano} and references therein) with nonextensive statistical mechanics. Indeed, the frequent occurrence in natural and artificial systems of the associated probability distributions would rely on this $q$-CLT, in the same way that the frequent occurrence of Gaussians relies on the standard $G$-CLT.
Further exploration of the $q$-CLT is in progress, addressing among other things (i) the effects of varying $p$ and of imposing the $q$-product elsewhere than on $r_{N,0}$; 
(ii) the results of extending the present procedure from $q_e \le 1$ to the entire region $q_e <3$. This extension will presumably require, for $1 \le q_e <3$, a new formula in place of Eq. (7). \\


Longstanding conversations on the subject of two of us (L.G.M. and C.T.)  with F. Baldovin, E.P. Borges and S.M.D. Queiros, and useful remarks from  J.D. Farmer, F. Lillo, S. Steinberg and H. Suyari are acknowledged. We have benefitted from partial financial support by Pronex/MCT, Faperj and CNPq (Brazil), and SI International and AFRL (USA).  M. G-M. was generously supported by the C.O.U.Q. Foundation and by Insight Venture Management. 

\end{document}